\documentstyle[12pt]{article}

\title{\bf $1/N_c$ expansion for the partition function\\ 
in four fermion models\footnote{Partially
supported by the Swiss National Foundation}}
\author{
\\
A.Barducci, R.Casalbuoni, M.Modugno, G.Pettini\\
{ \small Dipartimento di Fisica, Univ. di Firenze and
I.N.F.N., Sezione di Firenze }\\[0.5cm]
R.Gatto\\
{\small D\'epartement de Physique Th\'eorique, Univ. de Gen\`eve}
}\date{UGVA-DPT 1996/03-916}

\begin{document}
\maketitle

\begin{abstract}
\indent\noindent

We present a derivation of the bosonic contribution to the thermodynamical 
 potential of four fermion models by means of a 
$1/N_c$-expansion of the functional integral  defining the partition
function. This expansion turns out to be particularly useful 
to correct the mean field approximation
expecially  at low temperatures, where the relevant degrees of freedom
are low-mass bosonic excitations (pseudogoldstones).

\end{abstract}

\section{Introduction}

\indent\noindent

Four fermion models, like the NJL \cite{njl} and the 
Gross-Neveu model \cite{gn}, have been widely
studied, expecially in connection with chiral symmetry
breaking and restoration at finite densities and temperatures.
The thermodynamics of these models may offer useful hints 
in understanding the properties of thermal QCD,  
expecially those related to the chiral symmetry. 
Such a knowledge is expected to be important to
interpret heavy-ion collisions at relativistic energies
and for the description of early phases of cosmology.

Usually such studies are
carried out in the mean field approximation, 
for instance by using the Dolan-Jackiw formalism \cite{dolan}
of imaginary-time Green functions.

A review of the applications of the NJL model to 
hadronic and nuclear physics can be found in the report by
Vogl and Wise \cite{vogl}, to which we also refer for the 
vast literature on the subject. A calculation of the 
thermodynamical potential for the NJL model in the grand 
canonical ensemble, in the mean field approximation,
can be found for instance in ref.\cite{asakawa}.

The mean field approximation neglects fluctuations
around it, and it excludes in such a way possible
bosonic contributions to the thermodynamical potential. On 
the other hand such fluctuations, as they stand out from
possible resonant or non resonant bosonic scattering states,
can be expected to be of great importance in the physical problem.
For instance the pions, which are pseudogoldstones related to
spontaneous chiral symmetry breaking, are expected to
dominate the low energy properties of the system, as they 
correspond to the lowest mass bosonic excitations. We recall
that in non-relativistic statistical mechanics a resembling 
situation is the calculation of the second virial coefficient,
which is related to time-delay in the collision, and, in particular, 
bound states (two molecules combining to form a new
molecule) lead to a modification of the system pressure \cite{ma}.

Unfortunately, even in a simple field theoretic model, such as
NJL, exact analytical calculations appear unfeasible 
and one has to resort to approximate methods, such as large-N 
expansion. Such type of expansion has been used since a long 
time in statistical field theory. Well-known applications 
are to the $O(N)$ symmetric field theory 
\cite{zinn} and to the Gross-Neveu model \cite{gro}.

A calculation of fluctuations has been recently carried out 
\cite{klevansky} for the NJL model by evaluating the thermodynamical
potential through the method of coupling constant integration,
as developed in the book of Fetter and Walecka \cite{fetter},
by including terms to order $1/N_c$, where $N_c$ is the
number of colors ($N_c=3$ in QCD). The $1/N_c$ contributions
are taken into account through diagrammatic method. The use of
$1/N_c$ expansion goes back to Witten \cite{witten} and a 
review is given in the Erice lectures by Coleman \cite{coleman}.

The use of diagrammatic methods, besides its cumbersome
features, has also a delicate aspect, as it may easily 
destroy general symmetries of the system by improper selection
of the set of diagrams \cite{kadanoff}. It is therefore 
strongly preferable to recur to procedures which 
automatically guarantee conservation of the system symmetries.

In this note we shall develop a method based on the integral
functional formulation of field thermodynamics which
leads to straightforward calculation of the $1/N_c$ 
expansion of the partition function. Within this method
conservation of symmetries is consistently and
naturally guaranteed.

\section{$1/N_c$ expansion}
\indent\noindent

Let us consider a model in $D=d+1$ dimensions defined by the following 
four fermion lagrangian \cite{njl,gn}

\begin{equation}
{\cal L}=\bar{\psi}\left( i \hat{\partial}-M\right)\psi 
+ {g^2 \over2} \left[(\bar{\psi}\psi)^2 -
(\bar{\psi}\gamma_5\vec{\tau}\psi)^2\right]
\end{equation}
where $M$ is the bare fermion mass, 
$\tau_i$ are the matrices of the fundamental
representation of the flavor group $SU(N_f)$, and $\psi$ 
consists of $N_c$ separate
$N_f$-dimensional representations of $SU(N_f)$.

The generating functional (evaluated at vanishing sources) is
\begin{equation}
Z =\int{\cal D}(\bar{\psi}\psi)~e^{\displaystyle
i\int {d^D x}\cal{L}}\label{zetaferm}
\end{equation}

By using the identities
\begin{eqnarray}
&&\int{\cal D}(\sigma)~e^{\displaystyle-{i\over 2}\int {d^D x}
(\sigma-g{\bar\psi}\psi)^2}=cost\nonumber\\
&&\int{\cal D}(\vec{\pi})~e^{\displaystyle-{i\over 2}\int {d^D x}
(\vec{\pi}-i g{\bar\psi}\gamma_5\vec{\tau}\psi)^2}=cost'
\end{eqnarray}
and by inserting them in eq.(\ref{zetaferm}), we can write
\begin{equation}
Z =\int{\cal D}(\bar{\psi}\psi){\cal D}(\sigma){\cal D}(\vec{\pi})~
e^{\displaystyle i\int {d^D x}
\left[\bar{\psi}(i\hat{\partial}- M + g\sigma
+i g\gamma_5\vec{\tau}\cdot\vec{\pi})\psi -
{1 \over 2}(\sigma^2+\vec{\pi}^2)\right]}
\end{equation}

By carrying out the integral over the fermion fields, we obtain an effective
action in which only the bosonic degrees of freedom do appear
\begin{equation}
Z =\int{\cal D}(\sigma){\cal D}(\vec{\pi})~
e^{\displaystyle i S_{B}}\equiv ~ e^{\displaystyle
i W}
\end{equation}

By  redefining the $\sigma$ field with the following shift
\begin{equation}
g\sigma\rightarrow g\sigma+M
\end{equation}
we can write
\begin{equation}
S_{B}=\int {d^D x}\left[-{1\over 2}(\sigma^2+\vec{\pi}^2)
 (1+\delta Z_D) - {M\sigma\over g}\right]-i\log {\rm det}
(i\hat{\partial} +g\sigma+i g\gamma_5\vec{\tau}\cdot\vec{\pi})
\end{equation}
where $\delta Z_D$ is an counterterm coming from the 
renormalization of the fermion loop \cite{gn}. 
In NJL $\delta Z_4=0$, since the theory
is non-renormalizable. In  this case the regularization of infinities
is obtained by means of a momentum cut-off.
 
To simplify the notations we define
\begin{equation}
N\equiv N_f N_c \quad;\quad \lambda\equiv Ng^2
\quad;\quad
\alpha\equiv {M \over Ng^2}
\end{equation}
\begin{equation}
\phi\equiv(g\sigma,g\vec{\pi})\quad;\quad
a\equiv (1,i \gamma_5\vec{\tau})
\end{equation}

By using the formal identity
{\sl log  det =  tr  log},
and carrying out explicitly the trace over colour indices,
we can write
\begin{eqnarray}
S_{B} &=&\int {d^D x}\left[ -{N\over 2\lambda}\phi^2 (1+\delta Z_D)
- N\alpha\phi^j \delta_{j1}\right]
-iN_c {\rm Tr}\log(i\hat{\partial} +\phi\cdot a)
\nonumber\\
&\equiv&\int {d^D x}{\cal L}_B(\phi)\label{boson}
\end{eqnarray}

At this point, by following the method of ref.\cite{jackiw},
we couple an external constant source $J$ to the field $\phi(x)$,
and expand $\phi(x)$ around its vacuum expectation value $\bar{\varphi}$
\begin{eqnarray}
\phi(x)&=&\phi'(x)+\bar{\varphi}\label{phiprimo}\\
\bar{\varphi}[J]&=&\langle0^+|\phi(x)|0^- \rangle_J = 
{\delta W\over\delta J}
\label{barphi}
\end{eqnarray}

Since $J$ is constant, from the Lorentz invariance of the vacuum, it 
follows that $\bar{\varphi}$ is constant too. By supposing $\bar{\varphi}[J]$ 
to be invertible, we can define the {\sl effective action} 
$\Gamma[\bar{\varphi}]$ 
as the Legendre transform of $W[J]$
\begin{equation}
\Gamma[\bar{\varphi}] = W[J[\bar{\varphi}]] - 
\int {d^D x}J[\bar{\varphi}]\bar{\varphi} \equiv -{\cal V}
(\bar{\varphi})\int {d^D x}\label{gamma}
\end{equation}

 It can be easily
shown that, at vanishing sources, $\bar{\varphi}$ must satisfy the 
stationary equation
\begin{equation}
{\delta {\cal V}({\varphi})\over
\delta{\varphi}}\Bigg|_{{\varphi}=\bar{\varphi}}=0
\label{gapw}
\end{equation}

By using eq.(\ref{barphi}), $W[J]$ can be written as
\begin{equation}
W[J]\equiv W_{0}[J]+W_{1}[J]
\end{equation}
where
\begin{equation}
W_{0}[J]=\int d^{D}x\Big({\cal L}_{B}(\bar{\varphi})+J\bar{\varphi}\Big)
\end{equation}
and $W_1[J]$ satisfies the integral equation \cite{jackiw}
\begin{equation}
W_{1}[J]\equiv -i\log\int{\cal D}(\phi')
\exp\left[i\int{d^D x}\left({\cal L}_{B}^{(2)}(\phi',\bar{\varphi})-\phi'
{\delta W_1\over\delta\bar{\varphi}}\right)\right]\label{wuno}
\end{equation}
and ${\cal L}_{B}^{(2)}$ is obtained from the lagrangian 
${\cal L}_B(\phi'+\bar{\varphi})$ after having subtracted constant and
linear terms in $\phi'$.

By using eq.(\ref{phiprimo}), eq.(\ref{boson}) becomes
\begin{eqnarray}
S_{B} &=&\int {d^D x}\left[ -{N\over 2\lambda}(\phi'+\bar{\varphi})^2 
(1+\delta Z_D)
- N\alpha(\phi'+\bar{\varphi})^j \delta_{j1}\right]\nonumber\\
&&
-iN_c {\rm Tr}\log(i\hat{\partial} +\bar{\varphi}\cdot a+\phi'\cdot a)
\label{sbos}
\end{eqnarray}

We now define
\begin{equation}
(i\hat{\partial}+\bar{\varphi}\cdot a)G(x_1-x_2)=\delta^D (x_1-x_2)
\end{equation}

The last term in eq.(\ref{sbos}) becomes
\begin{equation}
-iN_c {\rm Tr}\log(i\hat{\partial} +\bar{\varphi}\cdot a+\phi'\cdot a)=
-iN_{c}{\rm Tr}\log(i\hat{\partial} +\bar{\varphi}\cdot a)
-iN_c{\rm Tr}\log\left(1+G\phi\cdot a\right)
\end{equation}

Thus
\begin{equation}
W_{0}[J]=\int {d^D x}\left[ -{N\over 2\lambda}\bar{\varphi}^2 
(1+\delta Z_D)
- N\alpha\bar{\varphi}^j \delta_{j1}+J\bar{\varphi}\right]
-iN_c {\rm Tr}\log(i\hat{\partial} +\bar{\varphi}\cdot a)
\end{equation}
and its final form, at $J=0$, is the standard mean field
fermionic term
\begin{equation}
W_{0}=\int {d^D x}\left[ -{N\over 2\lambda}\bar{\varphi}^2 
(1+\delta Z_D)
- N\alpha\bar{\varphi}^j \delta_{j1}+
iN{D\over 2}\int{d^D p\over (2\pi)^D}\log(p^2+\bar{\varphi}^2)\right]
\end{equation}
which is the only part which survives in the $N_c\rightarrow+\infty$
limit, since $W_1$ is of order $1/N_c$ with respect to $W_0$.

To evaluate $W_{1}$, one has to expand $S_B$ in eq.(\ref{sbos}) in powers
of $\phi'$, starting from the quadratic terms. Now, by ordering the terms
of the argument of the exponential in eq.(\ref{wuno}) in powers
of $1/N_c$ (instead of the $\hbar$ formal series of ref.\cite{jackiw}),
it turns out that the leading term is just the quadratic one (this can be
verified by the rescaling  $\phi'\rightarrow\phi'/\sqrt{N_c}$).

Thus, by keeping only the leading term of the $1/N_c$ expansion,
$W_{1}$ is given by
\begin{eqnarray}
e^{\displaystyle {iW_{1}}}&\equiv&  \int{\cal D}(\phi')
\exp\left\{-{iN\over 2\lambda}
\int{d^D x}\phi'^2 (1+\delta Z_D)\right.\nonumber\\
&&\qquad \left.-{N_{c}\over 2}
\int{d^D x_1}{d^D x_2}{\rm Tr}
\Big[G(x_1 - x_2)\phi'(x_1)\cdot a G(x_2-x_1)
\phi'(x_2)\cdot a\Big]\right\}
\end{eqnarray}

By Fourier transforming 
and evaluating the traces in the previous equation we finally obtain
\begin{equation}
W_1
= \sum_{j}{i\over2}\int{d^D x}\int{d^D p\over(2\pi)^D}
\log\left[{i N\over2\lambda}D^{-1}_{j}(p)\right]
\end{equation}
where $j$ labels each scalar/pseudoscalar 
field $\phi'_j$ and $D^{-1}_{j}(p)$ is given by
\begin{eqnarray}
D^{-1}_{j}(p)
&=& i\left[ (1+\delta Z_D) +i\Pi_{j}(p) \right]\nonumber\\
&=& i\left[ (1+\delta Z_D) - iD\lambda
\int{d^D q\over(2\pi)^D}\displaystyle{q\cdot(q+p)\pm \bar{\varphi}^2\over
\left[(p+q)^2 -\bar{\varphi}^2\right](q^2 - \bar{\varphi}^2)}
\right]\label{omega}
\end{eqnarray}
where $\pm$ refer to the scalar 
and pseudoscalar self energy $\Pi_{j}(p)$ respectively .

Notice that, apart from the linear term in $\bar{\varphi}_1$,
both $W_{0}$ and $W_{1}$ depend on the chirally invariant combination
${\bar{\varphi}}^2=g^2({\bar{\sigma}}^2+{\vec{\bar{\pi}}}^2)$, which
has to be determined by the eq.(\ref{gapw}).
From this condition it turns out that the physical value is
for zero pseudoscalar components, i.e. 
\begin{equation}
\bar{\varphi}=(g\bar{\sigma},\vec 0)
\end{equation}

\section{Free energy density at order $1/N_{c}$}
\indent\noindent

Taking into account the bosonic fluctuations is particularly
interesting also as far as the thermodynamics is concerned.
In fact, although depressed by a $1/N_{c}$ factor, pion
excitations should dominate the low-temperature equilibrium
properties of strong interacting matter (this is particularly
expected in the chiral limit, although there are limitations
in applying this method for zero current quark masses in low
dimensions, due to well known infrared problems).

The finite temperature extension of the formalism is
straightforward. In fact, for
imaginary times, one has simply to replace
\begin{equation}
  V_D\rightarrow-i\beta\int{d^d x}=-i\beta V_d\qquad ;\qquad
  \int {d^{D}p\over (2\pi)^{D}}
  \rightarrow{i\over\beta}\sum_{n=-\infty}^{n=+\infty}
\int{d^d p\over (2\pi)^d}
\end{equation}
and thus, at zero sources, the effective potential,
defined by the eq.(\ref{gamma}), is 
\begin{equation}
  {\cal V}_{\beta} =-i{W_{\beta}\over \beta V_d}
  =-{\log Z_{\beta}\over \beta V_d}
\end{equation}
which is the free energy density ${\cal F}$.

By means of the $1/N_c$ expansion
discussed in the previous paragraph, ${\cal F}$ 
can be separated in 
\begin{equation}
{\cal F} = {\cal F}^F + \sum_{j}{\cal F}^B_j
\label{state}
\end{equation}
with   
\begin{eqnarray}
{\cal F}^F &=& {N\over 2\lambda}\bar{\varphi}^2(1+\delta Z_D)
 + N\alpha \bar{\varphi}
-{ N D\over 2\beta}\sum_n
\int{d^d p\over (2\pi)^d}
\log\left(\bar{\varphi}^2+ \vec{p}^2+\omega^2_n\right)\nonumber\\
&=&{\cal V}_{0}(\bar{\varphi})-{ N D\over 2\beta}
\int{d^d p\over (2\pi)^d}\log\left(1+e^{\displaystyle 
-\beta\sqrt{\vec{p}^2+\bar{\varphi}^2} }
\right) \label{mean} \\
{\cal F}^B_j &=& {1\over2\beta}\sum_n\int{d^d p\over (2\pi)^d}
\log\left[{i N\over2\lambda}D^{-1}_j (p_{0_{n}},\vec{p})\right]
\label{free}
\end{eqnarray}
where the inverse bosonic propagator can be calculated 
at finite temperature by means of standard methods \cite{land}
\begin{eqnarray}
&&D^{-1} (p_{0_{n}},\vec{p})=D_0^{-1} (p_{0_{n}},\vec{p})-
iD\lambda\int{d^d q\over (2\pi)^d}
{\displaystyle n_{F}(E_q)\over E_q}\nonumber\\
&&\qquad+i{D\lambda\over4}\left(p^2-\epsilon^2_M\right)
\int{d^d q\over (2\pi)^d}
\left\{
{\displaystyle n_{F}(E_q)\over E_q}\left[{\displaystyle 1\over 
(E_q - p_{0_{n}})^2 - E_{q+p}^2}+
{\displaystyle 1\over 
(E_q + p_{0_{n}})^2 - E_{q+p}^2}\right]\right.\nonumber\\
&&\qquad+\left.
{\displaystyle n_{F}(E_{q+p})\over E_{q+p}}\left[{\displaystyle 1\over 
(E_{q+p} - p_{0_{n}})^2 - E_q^2}+
{\displaystyle 1\over 
(E_{q+p} + p_{0_{n}})^2 - E_q^2}\right]\right\}
\end{eqnarray}
where  
\begin{equation}
p_{0_{n}}=i\nu_n=2n\pi i/\beta\quad;\quad
n_{F}(E)={1\over e^{\displaystyle \beta E} +1}\quad;\quad
E_p=\sqrt{\vec{p}^2 + \bar{\varphi}^2 }
\end{equation}
and $\epsilon^2_M = 4\bar{\varphi}^2$ for the scalar
and $\epsilon^2_M = 0$ for the pseudoscalars.

The term ${\cal F}_F$ is the standard mean field
term coming from the one-loop calculations and it is purely fermionic,
whereas the second term, ${\cal F}_B$, is the contribution of bosonic 
fluctuations to the free energy. 
Expression (\ref{free}) and (\ref{mean}) put in eq.(\ref{state}) 
give the equation of state when evaluated
at the solution of the {\sl gap equation} (\ref{gapw})
at finite temperature, which is
\begin{equation}
{\delta {\cal F}(\varphi)\over\delta \varphi}
\Bigg|_{\varphi=\bar{\varphi}}=0
\label{gap}
\end{equation}

In particular, since the free energy ${\cal F}(\varphi)$ is a convex 
function of $\varphi$, $\bar{\varphi}$ must be the absolute 
minimum of ${\cal F}$ and this gives the evolution
of the fermion condensate with temperature.
The procedure adopted as well as the expressions obtained 
are general, although the bosonic part of the free energy
may be difficult to evaluate in the form of eq. (\ref{free}),
because of the sum over discrete energies $p_{0_{n}}$.
It is more convenient for calculations
to transform the sums into integrals by a two-steps standard
procedure (see for instance \cite{land}). The first step is to use
the residue theorem to trade the bosonic sums for an integral
over a circuit around the imaginary axis, and the second step
is rotating along a circuit around the real axis.

Namely one can use the relations ($n_{B}(z)=1/[\exp (\beta z)-1]$)
\begin{eqnarray}
{1\over\beta}\sum_n f(i\nu_n)&=&\lim_{\epsilon\rightarrow0}
{1\over2\pi i}\left[
\int_{-i\infty + \epsilon}^{+i\infty + \epsilon}\!\!\!\!\!\!\!
dz f(z)n_{B}(z)
+\int_{+i\infty - \epsilon}^{-i\infty - \epsilon}\!\!\!\!\!\!\!
dz f(z)n_{B}(z)\right]\nonumber\\
&=&\lim_{\epsilon\rightarrow0}
{1\over\pi i}\int_0^{+\infty}\!\!\!\!\!\!\!d\omega~
n_{B}(\omega)
\left[f(\omega+i\epsilon)-f(\omega-i\epsilon)\right]
+{1\over2\pi}\int_{-\infty}^{+\infty}\!\!\!\!\!\!\! d\omega f(i\omega)
\end{eqnarray}
provided $f(z)\exp(-\beta|z|)$ vanishes sufficiently 
fast at infinity, $f(z)$ has no singularities on the complex plane 
apart from a cut along the real axis, and $f(z)=f(-z)$ 
(as it turns out to be the case for the inverse 
propagator $i D^{-1}(z,\vec{p})$).

Therefore, the bosonic term can be cast in the form
\begin{eqnarray}
{\cal F}^B_j&=&\lim_{\epsilon\rightarrow0}{1\over2\pi i}
\int{d^d p\over (2\pi)^d}
\int_0^{+\infty}\!\!\!\!\!\!{d\omega}~
n_{B}(\omega)\log\left[{i D_j^{-1}(\omega+i\epsilon,
\vec{p})\over i 
D_j^{-1}(\omega-i\epsilon,\vec{p})}\right]\nonumber\\
&&+{1\over2\pi}\int{d^d p\over (2\pi)^d}\int_{0}^{+\infty}\!\!\!\!\!\! 
d\omega 
\log\left[i D_j^{-1}(i\omega)\right]
\end{eqnarray}
apart from infinities independent on $\beta$ and ${\bar{\varphi}}$.

Furthermore, we notice that the function $g(z)=i D^{-1}(z,\vec{p})$
satisfies the Schwarz reflection principle, $g^*(z)=g(z^*)$.
Thus, by reintroducing the explicit dependence on 
$\bar{\varphi}$ and $\beta$, we can write
\begin{eqnarray}
{\cal F}_j^B\left(\bar{\varphi},\beta\right) &=& \lim_{\epsilon\rightarrow0}
{1\over\pi}\int{d^d p\over (2\pi)^d}\int_0^{+\infty}\!\!\!\!\!\!{d\omega}~
n_{B}(\omega)\left({\rm arg}\left[
i D_j^{-1}(\omega+i\epsilon,\vec{p}; \bar{\varphi},\beta)\right]-\pi\right)
\nonumber\\
&&+{1\over2\pi}\int{d^d p\over (2\pi)^d}\int_{0}^{+\infty}\!\!\!\!\!\! d\omega 
\log\left[i D_j^{-1}(i\omega,\vec{p}; \bar{\varphi},\beta)\right]
\label{formula}
\end{eqnarray}
where $arg[f]$ is the argument $\theta\in[0,2\pi)$ of the complex number
$f\equiv|f|\exp(i\theta)$.

We notice that $(i)$ the second term of eq.(\ref{formula}) would be
the zero temperature term for a free boson gas, whereas now it is
temperature dependent since $i D^{-1}$ is a function of $\beta $;
$(ii)$ this term may need renormalization at $T=0$, due to momentum
divergencies \cite{root}\break (in $d=3$ divergent integrals are 
understood to be
regulated by a momentum cutoff). 
The first term, instead, is finite at any $T$, and tends to
zero for $T\rightarrow0$.

\section{Conclusions}
\indent\noindent

We have presented a functional integral procedure to calculate
the contribution of bosonic fluctuations to the partition
function in theories with four-fermion interactions by means
of a $1/N_{c}$ expansion. Such a procedure has the advantage,
as compared to diagrammatic methods, to automatically 
guarantee conservation of the system symmetries, through the
consistent use of the effective action formalism.

Our expression for the bosonic term in the free energy is given in 
eq.(37), as a sum of a finite term which vanishes at zero temperature
and and of a term which does not vanish in that limit, where
it must be regularized.

Fluctuations around the mean field are of greatest physical 
importance in connection to chiral symmetry breaking,
manifesting themselves through bosonic excitations,
some of which, by virtue of the Goldstone theorem,
correspond to the lowest bosonic mass excitations, 
and as such they dominate the thermodynamic behaviour at 
low temperatures where chiral symmetry is still far from being restored.

\vspace{1cm}
{\sl Note added}: Af\-ter com\-ple\-ting this work we ha\-ve seen a re\-cent 
pre\-print
by  E.N.Nikolov et al. [hep-ph/9602274] where the functional formalism is 
applied to the NJL model at zero temperature.

\end{document}